\documentclass[prd,
preprint,nofootinbib%
 ,secnumarabic%
,amssymb, amsmath,mathrsfs,nobibnotes,aps,11pt]{revtex4}
\usepackage{epsfig}
\usepackage{color}
\usepackage{ulem}
\usepackage{graphicx}
\usepackage{float}
\usepackage[caption=false]{subfig}
\usepackage{braket}
\usepackage{slashed}

\usepackage{tabularx}
\usepackage[colorlinks=true,citecolor=blue,linkcolor=blue]{hyperref}%
\def\bea {\begin{eqnarray}}
\def\eea {\end{eqnarray}}
\def\bc {\begin{center}}
\def\ec {\end{center}}
\def\nn {\nonumber}

\begin{document}

\title {Confronting nuclear equation of state in the presence of dark matter using GW170817 observation in relativistic mean field theory approach }

\author{Arpan Das$^{1,}$\footnote{ arpan@prl.res.in}}
\author{Tuhin Malik$^{2,}$\footnote{tuhin.malik@gmail.com}}
\author{Alekha C. Nayak$^{1,}$\footnote{ alekha@prl.res.in}}
\affiliation{$^1$Physical Research Laboratory, Ahmedabad 380009, India }
\affiliation{$^2$Department of Physics, BITS-Pilani, K.K. Birla Goa Campus, Goa 403726, India.}

\date{\today} 

\def\be{\begin{equation}}
\def\ee{\end{equation}}
\def\bearr{\begin{eqnarray}}
\def\eearr{\end{eqnarray}}
\def\zbf#1{{\bf {#1}}}
\def\bfm#1{\mbox{\boldmath $#1$}}
\def\hf{\frac{1}{2}}
\def\sl{\hspace{-0.15cm}/}
\def\omit#1{_{\!\rlap{$\scriptscriptstyle \backslash$}
{\scriptscriptstyle #1}}}
\def\vec#1{\mathchoice
        {\mbox{\boldmath $#1$}}
        {\mbox{\boldmath $#1$}}
        {\mbox{\boldmath $\scriptstyle #1$}}
        {\mbox{\boldmath $\scriptscriptstyle #1$}}
}

\begin{abstract}
We confront admixture of dark matter inside neutron star using gravitational wave constraints coming from binary neutron star merger.
We consider a relativistic mean field model including $\sigma-\omega-\rho$ meson
interaction with NL3 parameterization. We study fermionic dark matter interacting with nucleonic matter via
Higgs portal mechanism. We show that admixture of dark matter inside the neutron star soften the equation state and lower the value of tidal deformability.
 Gravitational wave GW170817 observation puts an upper bound on tidal deformability of
a binary neutron star with low spin prior at 90\% confidence level, which disfavors stiff equation of state such as Walecka model with NL3 parameterization.
However, we show that
 Walecka model with NL3 parameterization with a fermionic dark matter 
 component satisfy the tidal deformability bound coming from the GW170817 observation. 
\end{abstract}

\pacs{....}
\maketitle

\section{Introduction}

 Compact objects like neutron stars (NS) are nature's laboratory which can shed light directly or indirectly on the different branches of physics such as
 low energy 
 nuclear physics, QCD under extreme conditions, the general theory of relativity etc. A neutron star is one of the remnants of a dying star
 undergoing gravitational collapse. Gravitational collapse of stars with a mass range between 1.4-3.0$M_{\odot}$ evolve into a neutron star. 
 Neutron degeneracy pressure inside the neutron star makes it hydrostatically stable against the gravitational collapse. If the mass of a
 dying star is very large (beyond 10$M_{\odot}$) then the stellar remnant will overcome the neutron degeneracy pressure and gravitational
 collapse will produce a black hole. Matter density inside the neutron star can be  as large as few 
 times nuclear saturation density ($n_B=0.16$ fm$^{-3}$). Interior of a neutron star provides a situation to study the behavior of matter
 at extreme conditions. In this context, the possible equation of state (EoS) of infinite nuclear matter has been explored extensively
 (for a brief review see \cite{JMLattimer2013}). The main challenge in the description of matter at high densities inside neutron stars
 is to develop a model
 which  not only describes matter at high densities, but also the properties of matter observed at saturation densities \cite{Walecka1974,
 SerotWalecka1986,boguta}. Valid nuclear EoS has to satisfy presently well accepted empirical and experimental  constraints \cite{GambhirRing1990,Ring1996,dutra},
 e.g. ground state properties of 
 spherical and deformed nuclei, saturation density, binding energy, symmetry energy, compression modulus etc. as well 
 as constraints coming from infinite nuclear matter e.g., neutron star mass radius relation, tidal deformability etc. Rotating neutron stars 
 or pulsars give important information about superfluid nature of nucleon inside the neutron star \cite{BaymRuderman1969,Andersson2016}. 
 Superfluidity of nucleons is important to explain timing irregularities (glitch) of pulsars. From the high energy nuclear physics point of view, neutron star provides an ideal condition where high density QCD matter (quark matter phase, color superconducting phase etc.) can exist
 \cite{weber1999,Prakash2007,WeberMeixner2007,WeberNegreiros2010}. 
 Historically 
 neutron star mass radius relationship coming from solving
 Tolman-Oppenheimer-Volkoff (TOV) equation have been studied extensively to put constraints on the nuclear EoS  \cite{JMLattimer2013}. 
 However recent observation of gravitational waves from neutron star mergers  opens up another dimension in the study of the nuclear EoS\cite{Abbott2017a,Abbott2017}.

 On August 17, 2017, the Advanced LIGO and Virgo observatories detected the gravitational waves (GW) from a merging binary
 NS \cite{Abbott2017a}. GW170817 data open up a new way to understand the neutron star structure and the underlying EoS of dense matter.
 Details of the internal structure of the 
 neutron stars in the binary mergers become important as the orbital separation become comparable with the size of the bodies. For neutron star 
 binary merger the tidal field of the companion induces a  quadrupole moment to the other neutron star. The relation of the induced quadrupole
 moment to the external tidal field is proportional to the tidal deformability. Tidal deformability is sensitive to the mass, radius and tidal
 love number, which intern depends on the nuclear EoS. Using observed tidal deformability parameters of the neutron star merger one can put
 strong constraints on the neutron star EoS, for details and related studies see \cite{Hinderer2008,Hinderer2010,Postnikov2010, Yagi2013,Yagi2014,rezolla1,rezolla2}
 and references therein.
 The GW170817 observation puts an upper bound on tidal deformability of the combined binary NSs at 90\% confidence \cite{Abbott2017}. Consequently,
 this can be used to rule out certain equation of states of  neutron stars. 

  Observations of the kinematics of self gravitating objects such as galaxies and clusters of galaxies give strong hint of the existence of
 dark matter (DM). Cosmological observations tell us that this invisible matter cannot consist of baryons, it must be a new kind of 
 matter which interacts with the rest of the standard model particles very weakly. The exact nature of the dark matter, its coupling between
 standard model particles and the mass is still not known. However extensive studies on the particle physics dark matter models have put 
 strong constraints on the coupling constant and mass of the dark matter particles (for a recent review on dark matter physics see \cite{Plehn2017}).
 Among different proposals of dark matter, weakly interacting massive particle (WIMP) scenario has gained favor because
 it gives the correct prediction of the measured relic abundance of the dark matter today very naturally.

 Presence of dark matter inside  neutron star and its consequences have been discussed in the literature
 \cite{RajYu2018,GoldmanNussinov,GouldRoger,Kouvaris2007,KouvarisTinyakov2010,LavallazFairbairn2010,GuverErkoca2014,FanChang2012,
 PanotopoulosLopes2017,ellis,ellis2,Xiang2014}. This discussions include dark matter capture 
 by neutron star and heating of old neutron star in the galactic halo to a temperature detectable by upcoming infrared telescopes \cite{RajYu2018},
 trapped WIMPs inside the neutron star \cite{GoldmanNussinov}, charged massive dark matter particle and its effect on neutron star \cite{GouldRoger}, 
 heating of a neutron star due to dark matter annihilation \cite{Kouvaris2007,KouvarisTinyakov2010,LavallazFairbairn2010}, or the collapse of a 
 neutron star due to accretion of non annihilating dark matter \cite{GuverErkoca2014} etc. In Ref.\cite{ellis} authors have considered possible
 effects of a self interacting dark matter core on the maximum mass of a neutron star, mass-radius relation and on the NS 
 tidal deformability parameter. They have computed radial density and pressure profiles of the baryonic and dark matter components for different 
 nuclear EoS and different dark matter fractions. In Ref.\cite{PanotopoulosLopes2017}, the authors have considered  
 Walecka relativistic mean field model including $\sigma-\omega$ interaction for the nucleonic part \cite{Walecka1974,
 SerotWalecka1986,GambhirRing1990,Ring1996} and  fermionic dark matter inside the neutron star. Using mean field approximation they
 have calculated effective nucleon mass, variation of 
 $\sigma$ field, EoS  and the corresponding mass radius relation in this model. However, it is important to mention that the simple
 relativistic mean field model (RMF) model taken
 in this work is unsuccessful in producing nuclear saturation properties. This simple model is ruled out
 due to the fact that it gives high nuclear incompressibility ($\approx 500 $MeV) and very low nucleon mass \cite{ChungWang}.  
  Keeping this limitation of the simple $\sigma-\omega$ in this work we have considered the a generalized Lagrangian
 for the nucleonic sector including $\sigma-\omega-\rho$ meson interaction with NL3 parameterization \cite{Ring1997,boguta2}. The EoS of this model
 is disfavoured by GW170817 tidal deformability upper bound limit. However, we show that if we consider  fermionic dark matter  interaction
 via Higgs portal mechanism, then it can evade the GWs tidal deformability upper bound constraint. 
  We have also considered non gravitational interaction of the dark matter and the nucleon field  in a relativistic mean field approach.

 This paper is organized as follows: in Sec.\eqref{RMF} we discuss generalized Walecka model with NL3 parameterization.
 In Sec.\eqref{DM_sec}, we discuss fermionic dark matter model and its interaction with nuclear matter. The EoS of total
 Lagrangian density of dark matter and nuclear matter is presented in Sec.\eqref{Eos_sec}. Constraint from GWs tidal deformability
 on  EoS is discussed in Sec.\eqref{Tidal_sec}. Finally, we conclude in Sec.\eqref{Con}.

\section{Walecka model with NL3 parameterization}
\label{RMF}
In this section, we briefly summarize the relativistic mean field model (RMF) \cite{Walecka1974,
 SerotWalecka1986,GambhirRing1990,Ring1996}, which is also known as Quantum Hadron Dynamics
 (QHD) \cite{advance16}. In this framework, nucleons are quasiparticles with an effective medium dependent mass and baryon chemical potential.
 They move in the mean field of mesons. Simplest QHD model is known as $\sigma-\omega$ model. In this model nucleon-nucleon interaction is mediated by 
 the exchange of $\sigma$ and $\omega$ mesons. Properties of  symmetric nuclear matter has been studied in this model. In general
 $\sigma$ mesons give rise to a strong attractive central force and a spin-orbit nuclear force, on the other hand, $\omega$-mesons are 
 responsible for the repulsive central force. However, this simple model does not reproduce nuclear saturation properties, e.g. 
 compressibility \cite{ChungWang}.
 More advanced versions of QHD includes $\rho$ meson exchange interaction between 
 nucleons \cite{Ring1997}. Since protons and neutrons only differ in terms of their isospin projections,  $\rho$ mesons are included
 to distinguish between 
 these baryons and to give a better account of the symmetry energy. These vector mesons are charged, hence the reaction between  $\rho$ meson 
 and  proton will differ from the reaction between  $\rho$ meson and  neutron. In general one can also include photon field, however, neutron 
 star is assumed to be charge neutral, hence the contribution from the photon field can be neglected.

 Lagrangian including nucleon field, $\sigma$, $\omega$ and $\rho$ mesons and their interactions can be written as \cite{Ring1997},
\begin{eqnarray}
 \mathcal{L} = &\bar{\psi}\bigg[\gamma^{\mu}\left(i\partial_{\mu}-g_{v}V_{\mu}-g_{\rho}\vec{\tau}.\vec{b_{\mu}}\right)-(M_n+g_s\phi)\bigg]\psi
 +\frac{1}{2}\partial_{\mu}\phi\partial^{\mu}\phi-\frac{1}{2}m_s^2\phi^2-\frac{1}{3}g_2\phi^3-\frac{1}{4}g_3\phi^4\nonumber\\
 & -\frac{1}{4}V^{\mu\nu}V_{\mu\nu}+\frac{1}{2}m_V^2V^{\mu}V_{\mu}-\frac{1}{4}\vec{b^{\mu\nu}}.\vec{b_{\mu\nu}}+
 \frac{1}{2}m_{\rho}^2\vec{b^{\mu}}\vec{b_{\mu}}
 \label{Lag_m}
\end{eqnarray}
In general one can also include terms quartic in $\omega$ meson and $\omega-\rho$ interactions. However in the present work we have used NL3 
parameterization of the above Lagrangian. In this parameterization coupling of terms quartic in $\omega$ meson and $\omega-\rho$ interactions
are taken to be zero. In the above equation $\psi$ is  nucleon doublet, $\phi$, $V_{\mu}$ and $\vec{b_{\mu}}$ denotes $\sigma$, $\omega$, $\rho$ 
meson field respectively. $m_{s}$, $m_{V}$ and $m_{\rho}$ are the 
  masses of the mesons and $M_n$ denotes the nucleon mass. $g_s$, $g_v$ and $g_{\rho}$ are the scalar, vector and isovector coupling constants
  respectively. Field strength tensor of the vector and isovector mesons are given by, 
  \begin{equation}
   V_{\mu\nu}=\partial_{\mu}V_{\nu}-\partial_{\nu}V_{\mu},
  \end{equation}
  and,
  \begin{equation}
   \vec{b}_{\mu\nu}=\partial_{\mu}\vec{b}_{\nu}-\partial_{\nu}\vec{b}_{\mu}
  \end{equation}
  
  $\rho$ meson field can be written explicitly as, $\vec{b}^{\mu}=(b_1^{\mu},b_2^{\mu},b_3^{\mu})$. $b_3^{\mu}$ represents neutral $\rho^0$ meson and 
  $\rho^{\pm}$ are the orthogonal linear superposition of $b_1^{\mu}$ and $b_2^{\mu}$.
  \begin{eqnarray}
   b^{\mu}_{\pm}=\frac{1}{\sqrt{2}}(b_1^{\mu}\pm b_2^{\mu}).
  \end{eqnarray}
   $\vec{\tau}=(\tau_1,\tau_2,\tau_3)$ are the Pauli matrices, these are also the isospin operators. Proton and neutron are the
   different projections of nucleon in isospin space. Operation of $\tau_3$ on neutron and proton is as follows,
   \begin{equation}
    \tau_3|p\rangle=+1|p\rangle \qquad \& \quad \tau_3|n\rangle=-1|n\rangle
   \end{equation}
   Numerical values of the all the parameter of the 
 the Lagrangian are given in the following table\cite{Ring1997}, 
\begin{table}
\caption{Nucleon masses($M_n$), $\sigma$ meson mass($m_{s}$), $\omega$ meson mass($m_{v}$), $\rho$ mass$m_{\rho}$ and couplings $g_{s}$, $g_{v}$, $g_{\rho}$, $g_2$,  $g_3$ of NL3 parameterization }
 \begin{tabular}{|c|c|c|c|c|c|c|c|c|} 
 \hline
 $M_n$ & $m_{s}$ & $m_{v}$ & $m_{\rho}$ & $g_{s}$&  $g_{v}$& $g_{\rho}$&$g_2$ & $g_3$ \\
  (MeV) & (MeV) & (MeV) & (MeV) &  & & & $(fm^{-1})$ & \\
 \hline
 939 & 508.194 & 782.501 & 763.000 & 10.217 & 12.868 & 4.474& -10.431  & -28.885 \\
 \hline
 \end{tabular}
 \label{tab1}
\end{table}

\section{Interaction Lagrangian between nuclear matter and dark matter}
\label{DM_sec}
Due to the galaxy rotation, the compact object like neutron star pass through the dark matter halo and capture dark matter particle from it. Because of the high baryon density inside neutron star, DM particle loose energy due to its  interaction with neutrons. The strong gravitational force of the NS trap the DM after it loses some energy \cite{Xiang2014,goldman1989,Kouvaris2007}. 
There are also other mechanism such as conversion  neutrons to scalar dark matter, scalar DM production via bremsstrahlung increases the dark matter density inside the neutron star\cite{ellis,ellis2}. Since dark matter composed of 95 $\%$ total matter density, one could possible imagine that many compact objects composed of DM. The amount of dark matter inside NS also depend on the evolution history of NS, the environment where it lives. In Ref.\cite{brayeur}, the authors have shown that the binary neutron star systems might enhance DM accumulation probability inside NS.

 We consider fermionic dark matter  ($\chi$)  inside  the neutron star. 
 Here we consider the lightest neutralino which acts as a fermionic dark matter candidate \cite{PanotopoulosLopes2017,susy}. Dark matter is not directly
 coupled with the nucleons rather they are coupled to the Higgs field $h$. Coupling 
 between the dark matter and the Higgs field is $y$. For neutralino mass ($M_{\chi}$=200 GeV), the value of $y$
 varies from 0.001 - 0.1. We fix $y= 0.07$ for the rest of our analysis \cite{PanotopoulosLopes2017,susy}. The Higgs field is also coupled
 to the nucleons through 
 effective Yukawa coupling $\frac{fM_n}{v}$, where $f$ the proton-Higgs form factor and its value has been estimated to be approximately 0.35
 \cite{formfactor}.
 We have not considered $h^3$ and $h^4$ term in the Higgs potential, because in the 
 mean field approximation the value of the $h$ is small and the only dominant term is the $h^2$. So, the dark sector 
 Lagrangian and its interaction with the nucleons and Higgs field is given by \cite{PanotopoulosLopes2017}
 \begin{equation}
\mathcal{L}_{DM}=  \bar{\chi}\bigg[i\gamma^{\mu}\partial_{\mu}-M_{\chi}+yh\bigg]\chi+\frac{1}{2}\partial_{\mu}h\partial^{\mu}h-\frac{1}{2}M_{h}^2h^2
 +f\frac{M_n}{v}\bar{\psi}h\psi.
 \label{Lag_d}
 \end{equation}
 
 Direct detection experiment such as LUX \cite{lux} and XENON \cite{xenon} excluded dark matter-nucleon cross section above $\sim \, 8 \times 10 ^{-47}\, \text{cm}^{2}$  for dark matter mass range 30-50 GeV at 90 $\%$ C.L. . The invisible Higgs decay width tightly constraint the dark matter below $M_h/2$, hence the dark matter mass $m_{\chi}=200$ GeV evades these constraints. It is important to mention that dark matter may not be a single component, it may well be multicomponent system. Dark matter can be consist of low mass as well as high mass particles. As an example in Ref.\cite{PanotopoulosLopes2017} authors have considered heavy dark matter particles inside 
the neutron star.
 
 Here we have considered the average number density of nuclear matter is $10^3$ times larger than the  average dark matter density ($n_{DM}$), which gives the ratio between mass of the dark matter inside neutron star and mass of the neutron star to be $\sim \frac{1}{6}$
 \cite{PanotopoulosLopes2017}.
We know that nuclear saturation density $n_B\sim 0.16$fm$^{-3}$, so dark matter number density is $n_{DM}\sim 10^{-3}n_B\sim 0.16\times 10^{-3}$fm$^{-3}$.
 Number density of dark matter $n_{DM}=\frac{(k_{F}^{DM})^3}{3\pi^2}$, which gives $k_{F}^{DM}\sim 0.033$ GeV. In our calculations we have varied $k_{F}^{DM}$ from 0.02 GeV to 0.06 GeV. For 
 these values of $k_{F}^{DM}$ corresponding dark matter density will be different.

\section{Neutron star Equation of state and beta equilibrium}
\label{Eos_sec}
The Euler Lagrange Equation of motion for nucleon doublet ($\psi$), scalar($\phi$), vector($V^{\mu}$), isovector($\vec{b}^{\mu}$), DM particle ($\chi$)
and Higgs boson ($h$) can be derived from Lagrangian densities Eq.\eqref{Lag_m} and Eq.\eqref{Lag_d} as,
\begin{equation}
\begin{array}{r@{}l}
& \bigg[\gamma^{\mu}\left(i\partial_{\mu}-g_{v}V_{\mu}-g_{\rho}\vec{\tau}.\vec{b}_{\mu}\right)-\left(M_n+g_s\phi-\frac{fM_n}{v}h\right)\bigg]\psi {}=0, \vspace{5pt}\\
&\partial_{\mu}\partial^{\mu}\phi+m_s^2\phi+g_2\phi^2+g_3\phi^3+g_s\bar{\psi}\psi{}=0,  \vspace{5pt}\\
&\partial_{\mu}V^{\mu\nu}+m_{V}^2V^{\nu} {}=g_v\bar{\psi}\gamma^{\nu}\psi, \vspace{5pt} \\
&\partial_{\mu}\vec{b}^{\mu\nu}+m_{\rho}^2\vec{b}^{\nu} {}=g_{\rho}\bar{\psi}\gamma^{\nu}\vec{\tau}\psi,\vspace{5pt} \\
&\bigg(i\gamma_{\mu}\partial^{\mu}-M_{\chi}+yh \bigg)\chi {}=0, \vspace{5pt} \\
&\partial_{\mu}\partial^{\mu}h+M_{h}^2h {}=y\bar{\chi}\chi+\frac{fM_n}{v}\bar{\psi}\psi,
\label{eq:motion}
\end{array}
\end{equation}
respectively.  The DM particle 
mass and Higgs particle mass are denoted as $M_\chi$ and $M_h$, respectively. 
Applying standard relativistic mean field approximation we get,
\begin{equation}
\begin{array}{r@{}l} 
&\phi_0 {}=\frac{1}{m_s^2}\bigg(-g_s\langle\bar{\psi}\psi\rangle-g_2\phi_0^2-g_3\phi_0^3\bigg),\vspace{5pt} \\
&V_0 {}=\frac{g_v}{m_V^2}\langle\psi^{\dagger}\psi\rangle= \frac{g_v}{m_V^2}(\rho_p+\rho_n), \vspace{5pt} \\
&h_0 {}=\frac{y\langle\bar{\chi}\chi\rangle+f\frac{M_n}{v}\langle\bar{\psi}\psi\rangle}{M_h^2},\vspace{5pt}\\
&b_0 {}=\frac{g_{\rho}}{M_{\rho}^2}\langle{\psi^{\dagger}}\tau_3\psi\rangle=\frac{g_{\rho}}{M_{\rho}^2}(\rho_p-\rho_n),\vspace{5pt}\\
&\bigg(i\gamma^{\mu}\partial_{\mu}-g_v\gamma^0V_{0}-g_{\rho}\tau_3\gamma^0b_0-M_n^{\star}\bigg)\psi {}=0,\vspace{5pt} \\
&\bigg(i\gamma^{\mu}\partial_{\mu}-M^{\star}_{\chi} \bigg)\chi {}=0, 
\label{eq:mean}
\end{array}
\end{equation}
where $M_n^{\star}$ and $M_\chi^{\star}$ are nucleon and dark matter effective mass respectively. $\rho_p$ and $\rho_n$ are the densities of proton and neutron respectively. 
The effective mass of nucleon and dark matter can be given as,  
\begin{equation}
\begin{array}{r@{}l}
&M_n^{\star} {}= M_n+g_s\phi_0-\frac{fM_n}{v}h_0, \vspace{5pt}\\
 &M_{\chi}^{\star} {}=M_{\chi}-yh_0. 
\label{eq:m_star}
\end{array}
\end{equation}
The baryon density ($\rho$), scalar density ($\rho_s$) and dark matter scalar density ($\rho_s^{\rm DM}$)
are
\begin{equation}
\begin{array}{r@{}l}
&\rho {}= \langle\psi^{\dagger}\psi\rangle = \frac{\gamma}{(2 \pi)^3}\int_0^{k_F}d^3k, \vspace{5pt} \\
& \rho_s {}= \langle\bar{\psi}\psi\rangle = \frac{\gamma}{(2 \pi)^3}\int_0^{k_F}\frac{M_n^\star}{\sqrt{M_n^\star{^2}+ k^2}} d^3k, \vspace{5pt}\\
& \rho_s^{\rm DM} {}= \langle\bar{\chi}\chi\rangle =  \frac{\gamma}{(2 \pi)^3}\int_0^{k_F^{\rm DM}}\frac{M_\chi^\star}{\sqrt{M_\chi^\star{^2}+ k^2}} d^3k,
\label{eq:dense}
\end{array}
\end{equation}
where $k_F$ and $k_F^{\rm DM}$ are the Fermi momentum for the nucleonic matter and dark matter respectively. 
$\gamma$ is the spin degeneracy factor of nucleon and  $\gamma=2$ for neutron and proton individually. 

The masses for both nucleon and dark matter depends on baryon density for fixed values of
dark matter Fermi momentum $k_F^{\rm DM}$ and coupling constants. These masses and coupling values are discussed in Table.\eqref{tab1} and Sec.\eqref{DM_sec}. To get the density 
dependent profile for $M_n^{\star}$ and $M_\chi^{\star}$, one needs to solve numerically Eq. \eqref{eq:dense} together with
the field equations Eq. \eqref{eq:mean} in self consistent manner. 
In this work, we have taken the average dark matter number density approximately 1000 time smaller 
than the average neutron number density. This implies the dark matter mass fraction with respect to 
the neutron star mass is $\simeq 1/6$.
The expectation values of the energy-momentum tensor or the stress tensor provide the energy density and 
pressure of the system in static case i.e., the EoS, which is given by,
\bea 
\epsilon = \langle T^{00}\rangle ~~~~{\rm and}~~~~
P= \frac{1}{3} \langle T^{ii}\rangle .
\eea
The expression for the total energy density ($\epsilon$) and the pressure ($P$) can be
obtained by combining the Lagrangian density Eq.\eqref{Lag_m} and Eq.\eqref{Lag_d} :
\bea 
\epsilon = g_v V_0(\rho_p+\rho_n)+g_{\rho}b_0(\rho_p-\rho_n)+\frac{1}{\pi^2}
  \int_0^{k_p}dk k^2\sqrt{k^2+(M_n^{\star})^2}\nonumber\\
   +\frac{1}{\pi^2}\int_0^{k_n}dk k^2\sqrt{k^2+(M_n^{\star})^2}+\frac{1}{\pi^2}\int_0^{k_{F}^{DM}}dk k^2\sqrt{k^2+(M_{\chi}^{\star})^2}\nonumber\\
  +\frac{1}{2}m_s^2\phi_0^2+\frac{1}{3}g_2\phi_0^3+\frac{1}{4}g_3\phi_0^4-\frac{1}{2}m_V^2V_0^2-\frac{1}{2}m_{\rho}^2b_0^2+\frac{1}{2}M_{h}h_0^2. 
  \label{energydensity}
  \eea  
  \bea
P = \frac{1}{3\pi^2}\int_0^{k_p}\frac{k^4 dk}{\sqrt{k^2+(M_n^{\star})^2}} 
 +\frac{1}{3\pi^2}\int_0^{k_n}\frac{k^4 dk}{\sqrt{k^2+(M_n^{\star})^2}} + \frac{1}{3\pi^2}\int_0^{k_{F}^{DM}}\frac{k^4 dk}{\sqrt{k^2+(M_{\chi}^{\star})^2}}\nonumber\\
 -\frac{1}{2}m_s^2\phi_0^2-\frac{1}{3}g_2\phi_0^3-\frac{1}{4}g_3\phi_0^4+\frac{1}{2}m_V^2V_0^2+\frac{1}{2}m_{\rho}^2b_0^2-\frac{1}{2}M_{h}h_0^2,
 \label{pressure}
\eea
$\rho_n$ and $\rho_p$ are the neutron and proton number density with $k_F^n$ and $k_F^p$ are the corresponding 
Fermi momentum of neutron and proton, respectively. The number densities and corresponding Fermi momenta are equal 
for the symmetric nuclear matter. The matter inside neutron star mainly composed of neutrons. However, the neutron will 
eventually $\beta-$ decays as,
\bea 
n \rightarrow p + e^- + \bar{\nu}, \\
n + \nu \rightarrow p + e^- .
\eea 
To maintain the neutron star matter charge neutral, muons ($\mu$) will appear when the chemical potential 
of the electrons reaches the muon rest mass ($m_{\mu} = 106$ MeV). For a given baryon density ($\rho=\rho_n+\rho_p$), 
the charge neutrality is given as,
\bea 
\label{ch01}
\rho_p=\rho_e+\rho_{\mu}
\eea 
and the $\beta-$ equilibrium condition is given as,
\bea 
\label{ch02}
\mu_n=\mu_p+\mu_e ~~~~~~~~~~~~~~{\rm and}~~~~~~~~~~\mu_e=\mu_\mu
\eea 
Where, the chemical potentials $\mu_p$, $\mu_n$, $\mu_e$ and $\mu_\mu$ are given as,
\begin{eqnarray}
   \mu_p &=& \frac{\partial\epsilon}{\partial \rho_p}=g_vV_0+g_{\rho}b_0+\sqrt{k_p^2+(M_n^{\star})^2} \\
   \mu_n &=& \frac{\partial\epsilon}{\partial \rho_n}=g_vV_0-g_{\rho}b_0+\sqrt{k_n^2+(M_n^{\star})^2} \\
   \mu_e &=& \sqrt{k_e^2+m_e^2} \\
   \mu_{\mu} &=& \sqrt{k_{\mu}^2+m_{\mu}^2}
\end{eqnarray}
The particle fractions of neutrons and protons will depend on both charge neutrality and 
the $\beta-$ equilibrium condition as given above. The self consistent numerical solution
of Eq.\eqref{ch01} and Eq.\eqref{ch02} will set the fraction of neutron, proton, electron and muon number density for a given baryon
density. The total energy density and pressure of leptons are given as,
\bea 
\epsilon_l &=&\sum_{l=e,\mu}\frac{1}{\pi^2}\int_0^{k_l}  k^2 \sqrt{k^2+m_l^2} dk \\
P_l &=& \sum_{l=e,\mu}\frac{1}{3 \pi^2}\int_0^{k_l} \frac{k^4 dk}{\sqrt{k^2+m_l^2}}
\eea 
The total energy density and pressure for $\beta-$ equilibrated neutron star matter are
\bea 
\epsilon_{\rm NM}= \epsilon_l + \epsilon , \\
P_{\rm NM}=P_l + P .
\eea 
  
\begin{figure}[h]
  \centering
    \includegraphics[width=9.30cm]{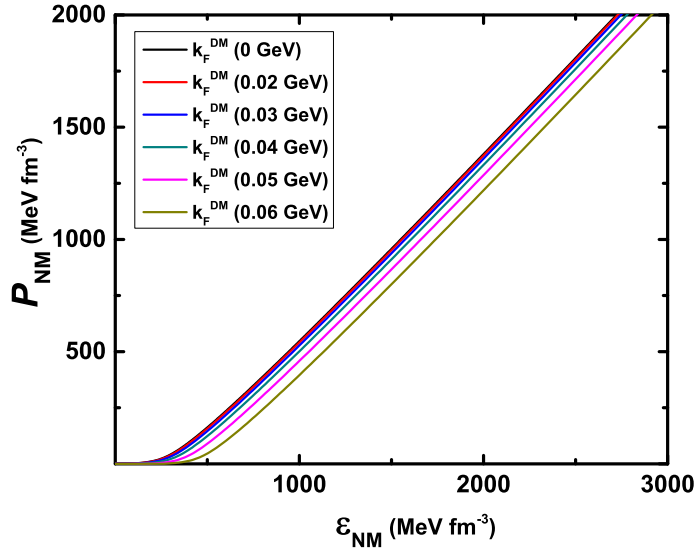}
    \caption{The equation of states of NS with different dark matter Fermi momenta $k_{F}^{DM}$, $0.0-0.06$ GeV with a step of 0.02 GeV. 
    The black line corresponds to EoS of NS without dark matter. The EoS becomes softer with increasing number density of dark matter inside the NS.}
    \label{Fig:eos}
  \end{figure}

   In Fig.\eqref{Fig:eos}, we plot pressure ($P_{\rm NM}$) as function of the total energy density ($\epsilon_{\rm NM}$) for different dark matter
   Fermi momentum $k_{F}^{DM}=0.0-0.06 $ GeV.  Fermi momentum $k_{F}^{DM}$=0.0 GeV corresponds equation of 
   state without dark matter. Increasing the value of  $k_{F}^{DM}$ from 0.02 GeV to 0.06 GeV, the EoS becomes softer,
   i.e. with increasing density of dark matter pressure reduces, which is consistent with earlier work \cite{PanotopoulosLopes2017}.
   This behaviour is evident from the expression of energy density and pressure from Eq.\eqref{energydensity} and Eq.\eqref{pressure},
   i.e. with increasing  $k_{F}^{DM}$
   the dark matter contribution in energy density increases much faster than the pressure.
   
\begin{figure}[h]
  \centering
    \includegraphics[width=9.30cm]{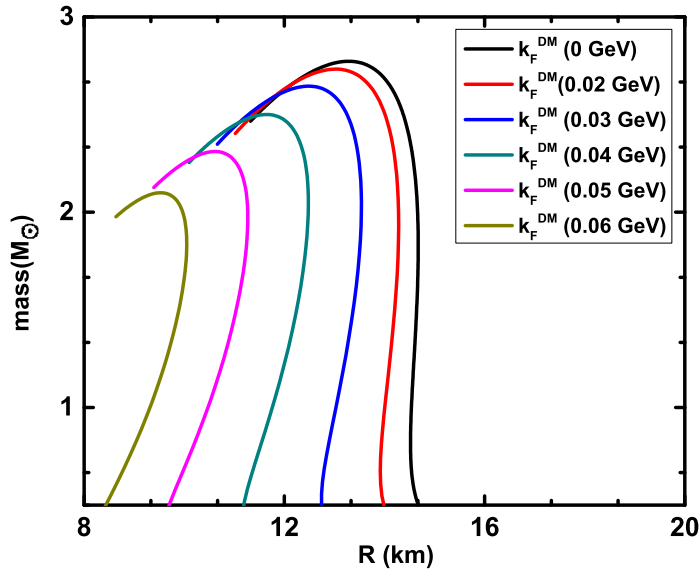}
    \caption{Mass radius relation of NS for different EoS with dark matter Fermi momentum $k_{F}^{DM}=0.0-0.06$ GeV.
    The maximum mass of NS without dark matter is  $2.8 M_{\odot}$
    and  $2.1 M_{\odot}$ with dark matter
   Fermi momentum $k_{F}^{DM}=0.06$GeV.}
    \label{fig_mr}
  \end{figure}

    Neutron star mass radius relationship can be obtained by solving TOV equation
    for a given nuclear matter EoS \cite{tov}. The EoS for the core is obtained from the Walecka Model with NL3 parameterization in the presence
    and absence of dark matter components. Crust EoS is modeled using the BPS EoS \cite{BPSeos} for the range of density
    $\rho \sim 4.8 \times 10^{-9}$ to $2.6\times 10^{-4}$ fm$^{-3}$.
    We use the polytropic form $P_{NM}(\epsilon_{NM})=a_1+a_2\epsilon_{NM}^{\gamma'}$ \cite{polytropicindex} to join the core and crust of the NS, 
    where $a_1$ and $a_2$ are
    obtained by matching the edge of the core at one end with the inner edge of the
    outer crust at other end, and $\gamma'$ is taken  $4/3$ \cite{tuhinprc}.    
    In Fig.\eqref{fig_mr} we plot the mass  radius of NS using the EoS as shown in Fig.\eqref{Fig:eos}.
    It is clear from the Fig.\eqref{Fig:eos} that the equation state is softer in the presence of larger dark matter density.
    A softer equation of state gives a lower value of the maximum mass of neutron star. From Fig.\eqref{fig_mr}, the maximum mass without
    dark matter ($k_{F}^{DM}=0$) $\sim$ $2.8 M_{\odot}$ and the value of the maximum mass of NS decreases with increasing dark matter density.
    
   In the context of dark matter inside the neutron star 
it is very natural to consider the formation of dark matter core like structure inside the neutron star with nonuniform distribution of the dark matter. This picture has been 
explored in various literatures\cite{ellis,Xiang2014}. We would like to point out that we are
focusing on the mainly qualitative aspect of the presence of dark matter
inside the neutron star. Effect of non uniformly distributed dark matter
inside the neutron star on its properties has been discussed in the Ref.\cite{Xiang2014}.
In the Ref.\cite{Xiang2014} authors have invoked two-fluid picture of the neutron star
containing nuclear matter fluid as well as dark matter fluid inside the neutron star.
 In that work, authors also have considered RMF picture for both nuclear fluid
as well as dark matter fluid, the
mass radius relation in the presence of dark matter for some specific EOS
is qualitatively similar to our result, e.g. in the
presence of dark matter, mass of neutron star can decrease. They have also shown 
that for mass of the dark matter ($M_{DM}$) can be large e.g. $0.33  M_{\odot}$. However, if one assumes RMF kind of situation for the dark matter sector then the coupling of the dark matter particle with a scalar, vector particles etc. will be free parameters of the theory. In the nuclear matter sector, the coupling
constants have been fixed keeping in mind the finite nuclei properties, but for the dark matter, these constraints are not available. Hence these free
coupling constants in the dark matter sector will make the model less predictive.

   Tidal deformability also depends upon the compactness and equation of state. In the next section, 
   we study the effect of dark matter on nuclear matter EoS  using the tidal deformability  constraint from  GW170817 observation.

\section{Tidal deformability constraint}            
\label{Tidal_sec} 
 One of the  important measurable structural properties of a binary merger is the 
tidal deformability. In a coalescing binary NS system, during the last stage of inspiral,  
{each NS} develops a mass {quadrupole} due to the tidal gravitational 
field induced by the other {NS forming} the binary. The tidal deformability describes 
the degree of deformation of a {NS} due to the tidal field of the 
companion {NS} and is sensitive to the nature of the EoS.
The tidal deformability is defined as, 
\begin{equation}
\label{eq2}
\lambda = \frac{2}{3}k_2R^5,
\end{equation}
where $R$ is the radius of the NS. The value of $k_2$ is typically in 
the range $\simeq 0.05-0.15$ \cite{Hinderer2008,Hinderer2010,Postnikov2010} for NSs and depends on the stellar 
structure. This quantity can be calculated using the following expression \cite{Hinderer2008}
\bea
&& k_2 = \frac{8C^5}{5}\left(1-2C\right)^2
\left[2+2C\left(y_R-1\right)-y_R\right]\nn\\
&&\times \bigg\{2C\left(6-3 y_R+3 C(5y_R-8)\right)\nn \\
&&+4C^3\left[13-11y_R+C(3 y_R-2)+2
C^2(1+y_R)\right] \nn \\
&& ~ ~
+3(1-2C)^2\left[2-y_R+2C(y_R-1)\right]\log\left(1-2C\right)\bigg\}^{-1},
\label{eq_k2}
\eea
where $C$ $(\equiv M/R)$ is the compactness parameter of the star of
mass $M$.  The quantity $y_R$ $(\equiv y(R))$ can be obtained by solving
the following differential equation
\bea
r \frac{d y(r)}{dr} + {y(r)}^2 + y(r) F(r) + r^2 Q(r) = 0
\label{eq_yr}
\eea
with
\bea
F(r) = \frac{r-4 \pi r^3 \left( \epsilon(r) - p(r)\right) }{r-2
M(r)}, \nonumber
\eea
\bea
\nonumber Q(r) &=& \frac{4 \pi r \left(5 \epsilon(r) +9 p(r) +
\frac{\epsilon(r) + p(r)}{\partial p(r)/\partial
\epsilon(r)} - \frac{6}{4 \pi r^2}\right)}{r-2M(r)} \\
&-&  4\left[\frac{M(r) + 4 \pi r^3
p(r)}{r^2\left(1-2M(r)/r\right)}\right]^2, \nonumber
\eea
along with TOV equation with proper boundary conditions \cite{tov,tuhin2}. One can then define the dimensionless tidal deformability: 
$\Lambda = \frac{2}{3}k_2 C^{-5}$.

 Individual dimensionless tidal deformability of two stars, $\Lambda_1$ and $\Lambda_2$
cannot be extracted independently from the observed gravitational
waveform. Instead, an effective dimensionless tidal deformability of the binary
$\tilde{\Lambda}$ can be extracted, which is a mass-weighted average of the individual dimensionless tidal deformability
 $\Lambda_1$ and  $\Lambda_2$. The effective tidal
deformability ($\tilde{\Lambda}$) of binary system in terms of $\Lambda_1$ and $\Lambda_2$ is defined as \cite{raithel}
\bea
\tilde{\Lambda}=\frac{16}{13}\frac{(m_1+12 m_2) m_1^4 \Lambda_1+(m_2+12m_1)m_2^4 \Lambda_2}{(m_1+m_2)^5}
\eea
where $m_1$ and $m_2$ are the masses of the neutron stars in the binary.
Similarly, masses of the two
companion neutron stars cannot be  measured directly, rather
the chirp mass, $\mathcal{M}_c=(m_1m_2)^{3/5}(m_1+m_2)^{-1/5}$, which is measured directly. By assuming low-spin
prior which is consistent with the binary neutron star systems
that have been observed in, GW170817 put an upper
bound on the NSs combined dimensionless tidal deformability and
 chirp mass with 90\% confidence \cite{Abbott2017}. This analysis predicts that the
combined dimensionless tidal deformability of the NS merger is
$\tilde{\Lambda}\leq 800$. It is important to note that a reanalysis of
GW170817 observation has been done assuming the
same EoS for both stars and this puts an upper limit on the dimensionless
tidal deformability, $\tilde{\Lambda}\leq 1000$ \cite{GWreanalysis}. However
lower bound on dimensionless tidal deformability can be put using the
investigation of the UV/optical/infrared
counterpart of GW170817 with kilonova models \cite{kilonova}. The lower bound of
dimensionless tidal  deformability is $\tilde{\Lambda}\geq 400$.

\begin{figure}[H]
  \centering
    \includegraphics[width=9.00cm]{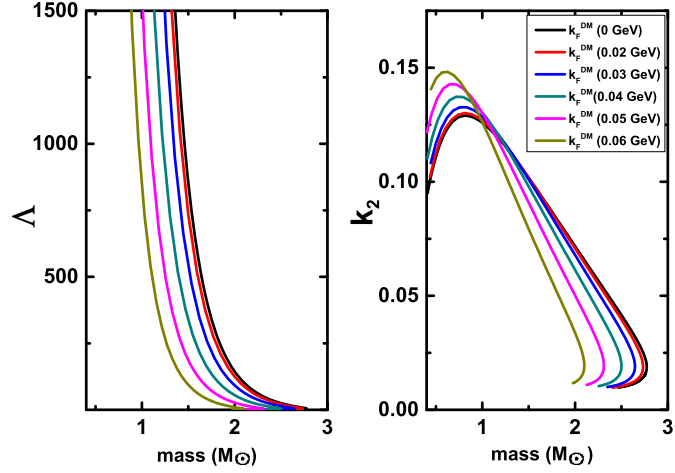}
    \caption{Dimensionless tidal deformability ($\Lambda$) of NS and Love number($k_2$) as a function of neutron star mass for different dark matter Fermi momentum is shown. }
    \label{fig_k2}
\end{figure}

One of the important goals of this work is to study the structural properties of neutron stars 
in the presence of the dark matter component. For the sake of arguments it is important to understand
the behavior of dimensionless tidal deformability and the tidal Love number of the 
companion neutron stars. Solving Eq.\eqref{eq_yr} and TOV equation with appropriate boundary conditions, we get $y_R$.
Using the value of $y_R$ and compactness one can get  $k_2$ using the expression Eq.\eqref{eq_k2}.  Left plot in Fig.\eqref{fig_k2}
shows the dimensionless tidal deformability $\Lambda$ and the right plot shows tidal Love number
$k_2$ as a function of the NS mass for our EoS with different dark matter density. The value of $k_2$ is of the range 0.09-0.13 
for a typical neutron star of mass 1.5$M_{\odot}$, which is expected \cite{Hinderer2008,Hinderer2010}. For a given neutron star mass
(say around 1.5$M_{\odot}$) EoS without the dark matter 
predicts larger radius and with increasing dark matter density radius decreases. Since the dimensionless tidal deformability is inversely 
proportional to the compactness $(C=M/R)$, its value is larger in the absence of dark matter. 

\begin{figure}[H]
  \centering
    \includegraphics[width=9.00cm]{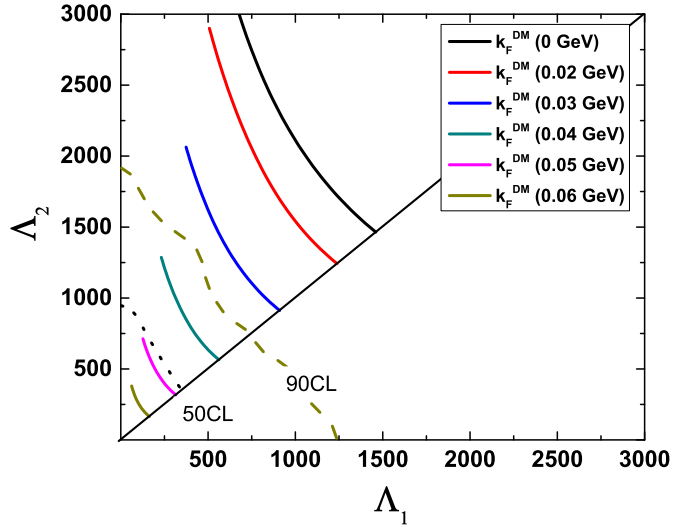}
    \caption{Tidal deformability parameters of the low and high mass components of binary neutron star merger (GW170817 observation).
              Dashed line and dotted line indicates 90\% and 50\% confidence limit for low spin priors  \cite{Abbott2017}. 
               The diagonal solid line corresponds to $\Lambda_1=\Lambda_2$ boundary. Walecka model with NL3 parameterization is disfavoured by GW 
               observation at 90\% C.L. in the absence of dark matter i.e. $k_{F}^{DM} = 0.0$ GeV. Note that low dark matter density,
               e.g. $k_{F}^{DM} = 0.03$ GeV is also disfavoured. However, Walecka model in NL3 parameterization with relatively higher DM density,
               e.g. $k_{F}^{DM} =0.04-0.06$ GeV is allowed by 90\% C.L.  of the GW170817 obsevation.}
    \label{fig_tidal}
\end{figure}

To study the tidal deformability constraint from the GW170817 observation on EoS of NS, we plot the combined tidal deformability of the 
binary system in
$\Lambda_1$, $\Lambda_2$ plane in Fig.\ref{fig_tidal}. $\Lambda_1$ and $\Lambda_2$ are the individual dimensionless tidal deformability
of the high mass $m_1$ and low mass
$m_2$ neutron stars in a binary, respectively. The curves are corresponding to the EoS with different dark matter density and
obtained by varying $m_1$ and $m_2$ independently. $m_1$ has been taken in the range $1.365<m_1/M_{\odot}<1.60$ and the range of
the $m_2$ is determined by 
keeping the chirp mass $\mathcal{M}_c$ fixed at 1.188$M_{\odot}$. The dashed and the
dot lines represent, respectively, the 90\% and 50\% confidence limits of the combined dimensionless tidal deformability
obtained from the GW170817 for the low spin prior. One can see from this plot that EoS given by the NL3 parameterization without 
dark matter component can be excluded at 90\% confidence level using the upper bound on tidal deformability of a binary system. However
if we consider dark matter component in neutron stars, then 
NL3 parameterized EoS comes within the 90\% confidence level. Hence a small component of dark matter inside a neutron star can revive 
well known EoS, which otherwise might be excluded by the GW170817 observation.

 \section{Conclusions}
 \label{Con}
      We have confronted the neutron star equation of state in the presence of dark matter 
      component using the gravitational wave constraint from the binary star merger. We have shown that for a uniformly
      distributed dark matter inside neutron star, the EoS becomes softer which eventually produces lower NS mass with 
      increasing dark matter density. We have taken Walecka model with NL3 parameterization in the nuclear matter sector. Walecka model
      with NL3 parameterization without dark matter admixture gives rise to a maximum mass of NS $\sim 2.8 M_{\odot}$. By increasing dark matter density 
      (Fermi momentum) inside neutron star reduces the value of maximum mass. Value of the maximum mass of neutron star changes from
      $2.8 M_{\odot}$ to $2.1 M_{\odot}$ by increasing dark matter Fermi momentum from $0.0$ GeV to $0.06$ GeV.
      One of the striking results of our analysis is that the  stiffer equation of states such as relativistic mean field model (Walecka model)
      with NL3 parameterization is ruled out at 90\% C.L. 
      using the GW170817 observation. However, in the presence of dark matter this constraint can be evaded and NL3 parameterization
      can be brought within  90\% C.L.. 
      
      \section*{Acknowledgements}
      T.M. acknowledges the hospitality provided by Physical Research Laboratory (PRL), Ahmedabad, India, during his visit, where the work has been initiated. 
      We would like to thank Prof. Hiranmaya Mishra for constant encouragement and support.

%\bibliography{i_ref}
%\bibliography{i_ref}

\end{document}